\documentclass[11pt,a4papers,floats]{article}
\usepackage{amsfonts,color,epsfig}
\usepackage{multirow}
\usepackage[multiple]{footmisc}
\usepackage{booktabs}
\newcommand{\nc}{\newcommand}

\newcommand{\rnc}{\renewcommand}
\renewcommand{\thefootnote}{\fnsymbol{footnote}}
\makeatletter
\rnc{\theequation}{\thesection.\arabic{equation}}
\@addtoreset{equation}{section}
\makeatother
\nc{\fig}[5]{
\begin{figure}[!htbp]
    \begin{center}
    \leavevmode
    \centerline{
        \includegraphics[width=#1, height=#2]{#3}
        }
    \caption[]{#4}
    \label{#5}
    \end{center}
\end{figure}}
\nc{\figs}[8]{
\begin{figure}[!htbp]
    \begin{center}
    \leavevmode
    \centerline{
        \includegraphics[width=#1, height=#2]{#3}
        \includegraphics[width=#4, height=#5]{#6}
        }
    \caption[]{#7}
    \label{#8}
    \end{center}
\end{figure}}

\begin{document}
\begin{flushright}
{arXiv:0812.2297v3 [hep-th]}\\
{\today}\\
\end{flushright}
\vspace{5mm}
\begin{center}
{{{\Large {\bf
The Role of Angular Momentum and Cosmic Censorship in the (2+1)-Dimensional Rotating
Shell Collapse}}}}\\[5mm]
{Robert B. Mann$^{a}$\footnote{email:rbmann@sciborg.uwaterloo.ca},
John J. Oh$^{b}$\footnote{email:johnoh@nims.re.kr},
and Mu-In Park$^{c}$\footnote{email:muinpark@gmail.com}}\\[5mm]

{\small \it {{${}^{a}$ {Department of Physics and Astronomy, University of Waterloo,
\\Waterloo, Ontario, N2L 3G1, Canada}\\[0pt]
${}^{b}$ Division of Interdisciplinary Mathematics, National Institute for
Mathematical Sciences, \\Daejeon, 304-350, Korea\\[0pt]
${}^{c}$ Research Institute of Physics and Chemistry, Chonbuk National University,\\
Chonju, 561-756, Korea\\[0pt]
}}}
\end{center}
\begin{abstract}{\small
We study the gravitational collapse problem of rotating shells in
three-dimensional Einstein gravity with and without a cosmological
constant. Taking the exterior and interior metrics to be those of
stationary metrics with asymptotically constant curvature, we solve
the equations of motion for the shells from the Darmois-Israel
junction conditions in the {\it co-rotating} frame. We study various
collapse scenarios with {\it arbitrary} angular momentum for a
variety of geometric configurations, including anti-de Sitter, de
Sitter, and flat spaces. We find that the collapsing shells can form
a BTZ black hole, a three-dimensional Kerr-dS spacetime, and an
horizonless geometry of point masses under certain initial
conditions. For pressureless dust shells, the curvature singularity
is {\it not} formed due to the angular momentum barrier near the
origin. However when the shell pressure is nonvanishing, we find
that for all  types of shells with polytropic-type equations of
state (including the perfect fluid and the generalized Chaplygin
gas), collapse to a naked singularity is {\it possible} under
generic initial conditions. We conclude that in three dimensions
angular momentum does not in general guard against violation of
cosmic censorship.}
\end{abstract}


\vspace{0.3cm}

\hspace{11.5cm}{Typeset Using \LaTeX}
\newpage
\renewcommand{\thefootnote}{\arabic{footnote}}
\setcounter{footnote}{0}

\section{Introduction}\label{sec:intro}

In  three-dimensional Einstein gravity, there are no dynamical
degrees of freedom, i.e., no gravitons that mediate the interactions
between massive objects in the vacuum \cite{Dese:84}, with or
without the cosmological constant, unless some higher derivative
terms, like the (gravitational) Chern-Simons term, are introduced
\cite{Dese:82}.   Although spacetime outside of matter is locally of
constant curvature, some non-trivial spacetime solutions to the
field equations exist that have  either black hole \cite{btz} or
cosmological event horizons \cite{Dese:84b,Park:98} when  the
cosmological constant is nonzero\footnote{In general cosmological
horizons in dS space do not necessarily imply the existence of black
hole horizons in the associated AdS space. However, in three
dimensions they are closely related, i.e, there is one-to-one
correspondence between the ``event'' horizons --  cosmological
horizons in dS and  outer black hole horizons in AdS -- for
overlapping ranges of ADM mass and angular momentum. See Ref.
\cite{Park:98} for the details of the mapping.}. Their metrics have
the general form
\begin{equation}
\label{eq:metbtz} (ds)^2 = -N^2(R) dT^2 + \frac{dR^2}{N^2(R)} +
R^2(d\varphi+N^{\varphi}(R)dT)^2
\end{equation}
with the lapse (squared) and shift functions
\begin{equation}
N^2(R) = \frac{(\alpha_{o} R^2 +\gamma_{o} )(\alpha_{i} R^2 +
\gamma_{i} )}{\ell^2R^2},~
{N^{\varphi}(R)=-\frac{\mbox{sign}(\alpha_{o} \alpha_{i})
\sqrt{\gamma_{o} \gamma_{i}}}{\ell R^2}}
\end{equation}
respectively. Here, $\alpha_{o/i}$ and $\gamma_{o/i}$ are constant
parameters whose signs (and values) depend on whether we are
considering the black hole solution in AdS space or the cosmological
solution in dS space.

The ADM mass and  angular momentum of this class of spacetimes, with
a cosmological constant $\Lambda = \pm 1/\ell^2$, are given by
\begin{equation}
\label{BTZ}
 {M_{BTZ}}=\frac{R_{o}^2+R_{i}^2}{ {8  G} \ell^2},~~{J_{BTZ}} =
\frac{2R_{o}R_{i}}{ {8 G} \ell}
\end{equation}
for the (BTZ) black hole solution (where $\alpha_o=\alpha_i=+1,~
\gamma_o=-R_o^2,~ \gamma_i=-R_i^2)$ \cite{btz} and
\begin{equation}
\label{KdS} {M_{KdS_3}}=\frac{R_{o}^2-R_{(i)}^2}{
{8  G} \ell^2},~~{J_{KdS_3}} =
\frac{2R_{o}R_{(i)}}{{8  G} \ell}
\end{equation}
for the (KdS$_3$) cosmological solution (where $\alpha_o=+1,~
\alpha_i=-1,~ \gamma_o=+R_o^2,~ \gamma_i=+R_{(i)}^2)$
\cite{Dese:84b,Park:98}, respectively\footnote{Here,  mass and
angular momentum agree with the definitions in Ref. \cite{Bala:01}
but differ in sign from Ref. \cite{Stro:01}.}, where the
three-dimensional Newton constant $G$ is assumed positive\footnote{
In three-dimensional gravity, there is no {\it a priori} reason to
fix the sign of Newton's constant. For some recent discussion  on
the sign choice, see Ref. \cite{Dese:82}.}.   For the black hole
case of Eq. (\ref{BTZ}), we have ${M_{BTZ}}\ge {J_{BTZ}}/\ell \ge 0
$ in order that there is no naked conical singularity and $R_{o/i}$
denotes the {\it outer/inner} horizon. On the other hand, for the
cosmological solution, there is no constraint on $M_{KdS_3}$ and
$J_{KdS_3}$ in order that the horizon exists, unless $J_{KdS_3}=0$
is considered \cite{Park:98}: Even $M_{KdS_3}<0$ is allowed also. In
this case, $R_{o}$ denotes the (cosmological) event horizon
\cite{Dese:84b} but $R_{(i)}$ does not signify an inner horizon; the
real parameter $R_{(i)}$ is introduced just for convenience
\cite{Park:98}.

Over the years black hole and cosmological solutions in three
dimensions have fascinated theorists because of the potential
insights they afford into quantum gravity. Amongst the most
intriguing applications \cite{mr,Pele:95,Ilha:99,Gutt:05} have been
black hole formation and the associated issues of gravitational
collapse and the cosmic censorship conjecture \cite{cch}. While
there are numerous examples of initial conditions that form a naked
singularity in the context of general relativity, none of them are
generic as required by the terms of cosmic censorship conjecture.
However it has been proposed that three dimensions might be an
exception \cite{Hube:04,mo}. A recent study of collapsing {\it shells}
in three dimensions with {\it no} angular momentum showed that a
naked singularity and a Cauchy horizon can form as the final result
of shell collapse for a {\it broad} variety of initial data
\cite{Pele:95,mo}. This feature is quite generic since the results
are independent of the types of  collapsing shell matter such as
pressureless dust, polytropic matter, and a generalized Chaplygin gas
(GCG).

Furthermore, similar behaviour also appeared in previous studies of
collapsing dust in three dimensions \cite{mr} and the formation of
topological black holes in four dimensions \cite{ms}. (See refs.
\cite{Ilha:99,Ilha:97} for higher-dimensional extensions.) The
formation of a naked singularity is of great importance in
association with the AdS/CFT correspondence
\cite{Hube:04,Hert:04,MyHub} since one might ask whether or not the
quantum correlation function at the AdS boundary is properly defined
in the presence of a bulk singularity. This in turn raises
additional issues associated with the collapse to a naked
singularity, particularly the proper inclusion of quantum (gravity)
effects in the bulk
\cite{Vaz:2007hn}.

Thus far the possible scenarios for the emergence of a naked
singularity have been explicitly analyzed for collapse to final
states (black holes or otherwise) with no angular momentum
\cite{mr,Pele:95,Ilha:99,Gutt:05,mo,ms,Ilha:97}. Indeed, relatively
little is explicitly known about the gravitational collapse of
matter with nonzero angular momentum in any dimension.
It is therefore natural to take into account an extension of
gravitational collapse to exterior black holes with rotation (or
other possible configurations) and to investigate how rotational
effects alter previous results  \cite{mr,mo} concerning   cosmic
censorship violation. While this problem is technically formidable
in higher dimensions, a study of  gravitational collapse of shells
in three dimensions is considerably more tractable. So we shall
consider this problem in our paper\footnote{
After completing this work, a paper \cite{Vaz:0805} investigating a
rotating dust {\it cloud} appeared. Their  ``no singularity'' result
is consistent with our result for dust shells in Sec. 3. However,
our results for shells with pressure in Sec. 4 imply that the naked
singularity would {\it not} be avoided even for rotating clouds with
pressure \cite{Mann:08}.}.

To this end, we introduce a co-rotating coordinate system on the
shell, which simplifies the matching procedure. The angular momentum
produces a potential barrier around the origin, preventing the shell
from contracting to zero size. We analyze the possible collapse
scenarios by investigating the effective potential  for all types of
equation of state. For a pressureless dust shell, we find that its
angular momentum prevents the creation of a singularity at the
origin, unlike the non-rotating case \cite{mo}. Previous work on
rotating dust shell collapse was carried out in the Hamiltonian
formalism, an alternative to the Darmois-Israel formalism
\cite{Cris:04}.  For a more restricted class of scenarios, they
found that singularities were also avoided.  Our work goes beyond
this insofar as we consider shells with pressure. In this case,
curvature singularities of finite spatial extent are formed before
they meet the barrier, and the effective potential and surface
stress-energy tensor diverge. Therefore under rather generic
conditions it is possible for a naked singularity to form, violating
cosmic censorship as in the non-rotating collapse scenario
\cite{mo}.

The outline of our paper is as follows. In Sec. \ref{sec:coll}, we
present a set-up for the two-dimensional hypersurfaces with
arbitrary rotations and a spherical symmetry. Considering a
co-rotating frame where the co-moving observer on the shell sees
only the radial motion yields a simplified equation of motion for
the shell. In Sec. \ref{sec:dust}, we consider the evolution of a
pressureless dust shell and compare it to the results for
non-rotating shell collapse \cite{mo}. In Sec. \ref{sec:pressure},
we consider  shells with pressure whose equations of state are the
polytrope type, which includes the perfect fluid and the GCG.
Finally, we shall summarize and discuss our results in Sec.
\ref{sec:discussion}.

\section{Set-up: Rotating Shells in Co-Rotating Frame}
\label{sec:coll}

In this section, we consider  rotating shells in three-dimensional
Einstein gravity with a cosmological constant $\Lambda$. The bulk
Einstein equation is given by
\begin{eqnarray}
G_{\mu \nu}+\Lambda g_{\mu \nu} = {8 \pi G}
T_{\mu \nu}
\end{eqnarray}
with a bulk stress-tensor $T_{\mu \nu}$. If we introduce a
two-dimensional hypersurface with surface stress-energy tensor
denoted by ${\mathcal S}_{ij}$, then the three-dimensional manifold
is divided into three parts --- the interior space ${\mathcal
V}_{-}$, the exterior space ${\mathcal V}_{+}$, and the thin-shell
hypersurface $\Sigma$. The metric away from the shell decomposes as
$g_{\mu\nu}=\Theta(\sigma)g_{\mu\nu}^{+} +
\Theta(-\sigma)g_{\mu\nu}^{-}$, where $\sigma$ is a geodesic
coordinate and $\Theta(\sigma)$ is the Heaviside step function
\footnote{$\Theta(\sigma)$ is equal to $+1$ if $\sigma>0$, $0$ if
$\sigma<0$, and indeterminate if $\sigma=0$. It has the following
properties : $\Theta^2(\sigma)=\Theta(\sigma)$,
$\Theta(\sigma)\Theta(-\sigma)=0$, and $d\Theta(\sigma)/d\sigma =
\delta(\sigma)$, where $\delta(\sigma)$ is the Dirac delta function.
 }. We shall use the coordinate system $(T,R,\phi)$ in the
interior and exterior spaces while we use the co-moving coordinate
system $(\tau,\phi)$ {on} the shell. Then, the evolution of the
shell is obtained by the Darmois-Israel matching conditions between
metrics and the corresponding extrinsic curvatures in the interior
and the exterior geometries \cite{israel},
 \begin{eqnarray}
 \label{eq:jc}
&&[g_{ij}]=0,\\
&& { 8 \pi G} {\mathcal
S}_{ij}=-([K_{ij}]-g_{ij}[K]) \label{eq:jc2}
\end{eqnarray}
where $[A]\equiv \lim_{\sigma\to 0} ({A_{+}}-{A_{-}})$ with the
subscripts `$+$' and `$-$' denoting exterior and interior
spacetimes, respectively. Greek letters $(\mu, \nu, \cdots)$ denote
three-dimensional spacetime indices, whereas Roman letters $(i, j,
\cdots)$ denote the two-dimensional indices on the shell. The
combination of the metric junction condition and the induced
Einstein's equation on the shell will describe the effective motion
of the shell.

If we take a co-rotating frame on the shell by introducing $d\varphi
= d\phi + \epsilon\frac{R_{o}R_{i}}{\ell {\mathcal R}^2(T)}dT$,
then the metric (\ref{eq:metbtz}) becomes
\begin{equation}
(ds)^2 = - N^2dT^2 + \frac{dR^2}{N^2} +
R^2\left[d\phi+\frac{\epsilon
R_{o}R_{i}}{\ell}\left(\frac{1}{{\mathcal R}^2(T)} -
\frac{1}{R^2}\right)dT\right]^2\label{eq:indmet}
\end{equation}
and each metric in both regions is simply expressed as
\begin{equation}
(ds)^2_{{\mathcal V}_{\pm}} = - N^2_{\pm}dT^2 +
\frac{dR^2}{N^2_{\pm}} + R^2\left[d\phi -
{N_{\pm}^{\phi} R^2} \left(\frac{1}{{\mathcal R}^2(T)}
- \frac{1}{R^2}\right)dT\right]^2, \label{eq:indmet2}
\end{equation}
with
\begin{equation}
N_{\pm}^2 = \frac{(\alpha_{o}^{\pm}R^2 +
\gamma_{o}^{\pm})(\alpha_{i}^{\pm}R^2 +
\gamma_{i}^{\pm})}{\ell^2R^2},~
{N_{\pm}^{\phi}=-\frac{\epsilon_{\pm} \sqrt{\gamma_{o}^{\pm}
\gamma_{i}^{\pm}}}{\ell R^2}}
\end{equation}
and $\epsilon_{\pm} \equiv {\rm sign}(\alpha_{o}^{\pm}
\alpha_{i}^{\pm})$. Here $\alpha^{\pm}_{o/i}$ and
$\gamma^{\pm}_{o/i}$ depend on the choice of spacetime. The black
hole and cosmological solutions are respectively given by
$\alpha_{o/i}=+1, \gamma_{o/i} <0$ and $\alpha_{o}=-\alpha_{i}=+1,
\gamma_{o}>0$, as explained in Sec. 1. Note that $T$ is a time
(space) -like coordinate and $R$ is a space (time) -like coordinate
in the exterior (interior) of an outer black hole horizon or
interior (exterior) of a cosmological horizon. Moreover, the
horizonless geometry with point masses, where $T$ is time-like and
$R$ is space-like always, corresponds to setting $\gamma_{o/i}>0$
with $\alpha_{o/i}=+1$ for AdS space and $\alpha_{{i}}=0,
\alpha_{{o}}=+1$ for  flat space \cite{Carl:98}.

On the shell's {surface $\Sigma$} with $T={\cal
T}(\tau)$ and $R={\mathcal R}{\cal T}(\tau)= {\mathcal R}(\tau)$, the metric {reduces
to}\footnote{ { Here we consider the same cosmological constant
parameter $\ell$ for both interior and exterior spacetimes. For the
different parameters $\ell_{\pm}$, the analysis is the same with the
rescaled interior and exterior coordinates $\tilde{R}_{\pm}=R_{\pm}
\ell/ \ell_{\pm}={\cal R}(\tau) \ell/ \ell_{\pm}$, without changing
the physical parameters $\alpha^{\pm}$ and $\gamma^{\pm}$. }}
\begin{eqnarray}
(ds)^2_{\Sigma} &=& - {\cal N}_{\pm}^2 d {\cal
T}^2(\tau) + \frac{d{\mathcal R}^2(\tau)}{{\cal N}_{\pm}^2}
+ {\mathcal R}^2 d\phi^2\nonumber\\
&=&g_{ij} dx^i dx^j \equiv -d\tau^2 + r_0^2 a^2(\tau) d\phi^2,
\end{eqnarray}
which yields (from the first junction condition in Eq. (\ref{eq:jc}))
\begin{eqnarray}
\label{eq:metrel} {\mathcal N}_{\pm}^4 \left(\frac{d{\cal
T}}{d\tau}\right)^2 = \left(\frac{d{\mathcal R}}{d\tau}\right)^2 +
{\mathcal N}_{\pm}^2,~~{\mathcal R}^2 = r_0^2 a^2(\tau).
\end{eqnarray}
The induced basis vectors and the normal vectors on $\Sigma$ are
\begin{equation}
\label{eq:basis} e_{\tau}^{\mu}=\left(\frac{d{\cal T}}{d\tau},
\frac{d{\mathcal R}} {d\tau}, 0\right),~~{e_{\phi}^{\mu} =\left(0,
0, 1\right),}
\end{equation}
and
\begin{equation}
n_{\mu} = \left(-\frac{d{\mathcal R}} {d\tau}, \frac{d{\cal
T}}{d\tau}, 0\right),
\end{equation}
respectively. Then, the non-vanishing components of the extrinsic
curvature defined by $K_{ij}=-n_{\alpha}(\partial_{j}e_{i}^{\alpha}
+ \Gamma_{\mu\nu}^{\alpha}e^{\mu}_{i}e^{\nu}_{j})$ are computed to
be
\begin{eqnarray}
K_{\tau\tau}^{\pm} = - \frac{d}{d\mathcal R}
\sqrt{\left(\frac{d{\mathcal R}} {d\tau}\right)^2 + {\mathcal
N}_{\pm}^2}, ~~{K_{\tau\phi}^{\pm} = \frac{{\rm
sign}(\alpha_{o}^{\pm} \alpha_{i}^{\pm})
\sqrt{\gamma_{o}^{\pm}\gamma_{i}^{\pm}}} {\ell {{\mathcal
R}}}},~~K_{\phi\phi}^{\pm} = {\mathcal R}
\sqrt{\left(\frac{d{\mathcal R}}{d\tau}\right)^2 + {\mathcal
N}_{\pm}^2}
\end{eqnarray}
and its trace is
\begin{eqnarray}
 K &\equiv& g^{ij}K_{ij} \nonumber \\
 &=&\frac{d}{d\mathcal R}
\sqrt{\left(\frac{d{\mathcal R}} {d\tau}\right)^2 + {\mathcal
N}_{\pm}^2} + \frac{1}{\mathcal R} \sqrt{\left(\frac{d{\mathcal
R}}{d\tau}\right)^2 + {\mathcal N}_{\pm}^2}.
\end{eqnarray}
Note that $K_{\tau\phi}^{\pm}$ does not vanish even though there is
no $(T,\phi)$ component of the metric on the shell where $R={\cal
R}(\tau)$, i.e., $f(R) \equiv -N^{\phi} R^2 ({1}/{{\mathcal
R}^2(\tau)} - {1}/{R^2})=0$. This is basically because
$\Gamma^{R}_{\tau \phi}$ has a term $\partial_R f(R)$ which does not
vanish on the shell.

We shall assume that the surface stress-energy tensor of the shell
is that of a perfect fluid
\begin{eqnarray}
{\mathcal S}_{ij} =(\rho + p) u_{i}u_{j}+p g_{ij},
\end{eqnarray}
 where $\rho$ is
an energy density, $p$ is a pressure, and $u^{i}$ is the shell's
two-velocity. Then, from the second junction condition
(\ref{eq:jc2}), the surface stress-energy tensor {with
three-velocity $u^{\mu}=(1,0,0)$} is straightforwardly evaluated to
be
\begin{eqnarray}
-\rho &=&{{\mathcal S}^{\tau}}_{\tau} = \frac{1}{
{8 \pi G}  \mathcal
R}(\beta_{+}-\beta_{-}), \nonumber \\
p&=&{{\mathcal S}^{\phi}}_{\phi} = \frac{d}{ { 8
\pi G} d{\mathcal R}}(\beta_{+}-\beta_{-}), \label{eq:eqns}
\end{eqnarray}
where $\beta_{\pm} \equiv \sqrt{(d{\mathcal R}/d\tau)^2+{\mathcal
N}_{\pm}^2}$. In addition, 
we have ${\mathcal S}_{\tau\phi}={(\rho+p)
u_\tau u_\phi}=0$ since ${u_{\phi}}$ vanishes on the shell in the
co-rotating frame. Hence $[K_{\tau\phi}]=0$, which gives
\begin{equation}
K_{\tau\phi}^{+} = {\frac{{\rm sign}(\alpha_{o}^{+}
\alpha_{i}^{+})\sqrt{\gamma_{o}^{+}\gamma_{i}^{+}}}{\ell {\mathcal
R}} = \frac{{\rm sign}(\alpha_{o}^{-} \alpha_{i}^{-})
\sqrt{\gamma_{o}^{-}\gamma_{i}^{-}}}{\ell {\mathcal R}} }=
K_{\tau\phi}^{-}, \label{eq:jpjm}
\end{equation}
implying $J_{+} = J_{-}$ on the shell. {Combining} both equations in
 (\ref{eq:eqns}) yields
\begin{eqnarray}
&& {\beta_{+}-\beta_{-} +{  8 \pi G} \rho
\mathcal R = 0},
\label{eq:eqn1}\\
&& {\frac{d}{d\mathcal R} (\rho {\mathcal R}) + {
p} = 0.} \label{eq:eqn2}
\end{eqnarray}
The first equation states that the (relativistic) energy of the
shell, $2 \pi \rho {\cal R}$ is balanced by the respective
difference of energies  $\beta_{\mp}/4G$, as measured from the
interior and exterior spacetimes with given gravitational
backgrounds. The second equation occurs because the shell is a
closed, adiabatic system, with the loss of shell energy $d( 2 \pi
\rho {\cal R})$ under expansion occurring at the expense of the work
$2\pi  p d {\cal R}$ done by the shell.

\section{Pressureless Dust Shells}\label{sec:dust}

For a pressureless ($p=0$) dust shell, Eq. (\ref{eq:eqn2}) yields
$\rho=m_{0}/{2 \pi \mathcal R}$, where $m_{0}$ is an initial rest
mass of the shell and is assumed to be  a non-vanishing positive
constant.  Inserting this into Eq. (\ref{eq:eqn1}) gives {the}
equation of motion for the shell
\begin{equation}
\label{Junction1} \sqrt{\left(\frac{d{\mathcal R}}{d\tau}\right)^2 +
{\mathcal N}_{+}^2} -\sqrt{\left(\frac{d{\mathcal
R}}{d\tau}\right)^2 + {\mathcal N}_{-}^2} + 4 G m_{0} =0 ,
\end{equation}
or alternatively
\begin{equation}
\left(\frac{d\mathcal R}{d\tau}\right)^2 + V_{{\rm eff}}(\mathcal R)
= 0,\label{eq:eqx}
\end{equation}
where the effective potential is
\begin{equation}\label{veff1}
V_{{\rm eff}}(\mathcal R) = {-\frac{1}{4(4 G m_{0})^2}} \left[(4 G
m_{0})^4 -2({\mathcal N}_{+}^2+{\mathcal N}_{-}^2)(4 G m_{0})^2 +
({\mathcal N}_{+}^2-{\mathcal N}_{-}^2)^2\right].
\end{equation}
 Note that this equation describes the one-dimensional {Newtonian} motion
of a point particle with zero energy in the potential $V_{{\rm
eff}}(\mathcal R)$ insofar as surfaces of constant $\tau$ are
spacelike\footnote{In the opposite case, when surfaces of constant
${\cal R}$ are spacelike, the system corresponds to a
non-conservative system with a time-dependent potential $V_{{\rm
eff}}({\cal R})^{-1}$, due to absence of a time-like Killing vector.
This situation is analogous to that of the $S_0$-brane geometry
\cite{Gutp:02}. For a related analysis in a $S_0$-brane geometry,
see Ref. \cite{Burg:03}.}. The behaviour of the shell will depend on
the number of roots of the effective potential in $\mathcal R>0$
\cite{mo}.

Now if we define the dimensionless parameters (`$-$' for AdS and `$+$' for dS)
\begin{eqnarray}
x\equiv {\mathcal
R}/\ell,~ t\equiv \tau/\ell,~{k}_{o/i}
\equiv \gamma_{o/i}/\ell^2 =\mp ({x_{o/i}})^2,
\end{eqnarray}
with the overdot denoting $\partial /\partial t$, then Eq. (\ref{eq:eqx})
becomes
\begin{equation}
\dot{x}^2 + V_{{\rm eff}}(x)=0, \label{eq:eqnmot}
\end{equation}
where the effective potential is
\begin{equation}
\label{V_eff:p=o} V_{{\rm eff}}(x)=\frac{1}{m_{0}^2 x^2} (a_8 x^6 +
a_6 x^4 + a_4 x^2 + a_2)
\end{equation}
and we employ henceforth  the convention $8 G \equiv 1$ for
convenience, unless otherwise stated\footnote{This differs from that
of Ref. \cite{mo} by $\pi$.}. The coefficients are
\begin{eqnarray}
&& a_{8} = - (\alpha_o^{+}\alpha_{i}^{+}-\alpha_{o}^{-}\alpha_{i}^{-})^2,\nonumber\\
&& a_{6} = 2\left[ (\alpha_o^{+}\alpha_{i}^{+}
+\alpha_{o}^{-}\alpha_{i}^{-})m_{0}^2/4 -
(\alpha_{o}^{+}\alpha_{i}^{+}-\alpha_{o}^{-}\alpha_{i}^{-})
(\alpha_{o}^{+}k_{i}^{+}+\alpha_{i}^{+}k_{o}^{+}-\alpha_{i}^{-}k_{o}^{-}-\alpha_{o}^{-}k_{i}^{-})\right]\nonumber\\
&& a_{4} = - \left[m_{0}^2/4 - (\alpha_{o}^{+}k_{i}^{+}+\alpha_{i}^{+}k_{o}^{+}+\alpha_{i}^{-}k_{o}^{-}+\alpha_{o}^{-}k_{i}^{-})\right]^2  \nonumber\\
&& \qquad\qquad + 4(\alpha_{o}^{+}\alpha_{o}^{-}k_{i}^{+}k_{i}^{-}+\alpha_{i}^{+}\alpha_{i}^{-}k_{o}^{+}k_{o}^{-} + \alpha_{o}^{+}\alpha_{i}^{-}k_{i}^{+}k_{o}^{-}+\alpha_{o}^{-}\alpha_{i}^{+}k_{o}^{+}k_{i}^{-}),\nonumber\\
&&a_{2} = k_o^{+}k_{i}^{+} m_{0}^2=
k_{o}^{-}k_{i}^{-}m_{0}^2\nonumber
\end{eqnarray}
with the condition $k_{o}^{+}k_{i}^{+} = k_{o}^{-}k_{i}^{-}$
($J_{+}=J_{-}$) on the shell from Eq. (\ref{eq:jpjm}). Note that a
non-vanishing coefficient $a_8$ implies the interior and exterior
geometries differ; for example  exterior AdS and interior dS
spacetimes.

The coefficient $a_2$, which is positive, is peculiar to geometries
with rotation; if there is no rotation then $a_2$ vanishes
($k_{i}^{\pm}=0$), and the effective potential agrees with that
for the non-rotating case \cite{mo}.

The effective potential can be classified by the values of its
coefficients. For a non-vanishing $a_8$ (different geometries), the
effective potential behaves as $V_{{\rm eff}}(x) \approx a_2/
m_{0}^2 x^2 \rightarrow +\infty$ as ${x\rightarrow 0}$ corresponding
to a centrifugal barrier around the origin at $x=0$. Hence the shell
cannot collapse to zero size. We also have the asymptotic behaviour
$V_{{\rm eff}}(x) \approx a_8 x^4/m_{0}^2 \rightarrow -\infty$ as
$x\rightarrow \infty$ since $a_8 <0$.  The shape of the effective potential is one of four types, depending on the values of the parameters (as illustrated in Fig.
\ref{fig:efpa8}) and its numerator is a cubic polynomial in  $x^2$.

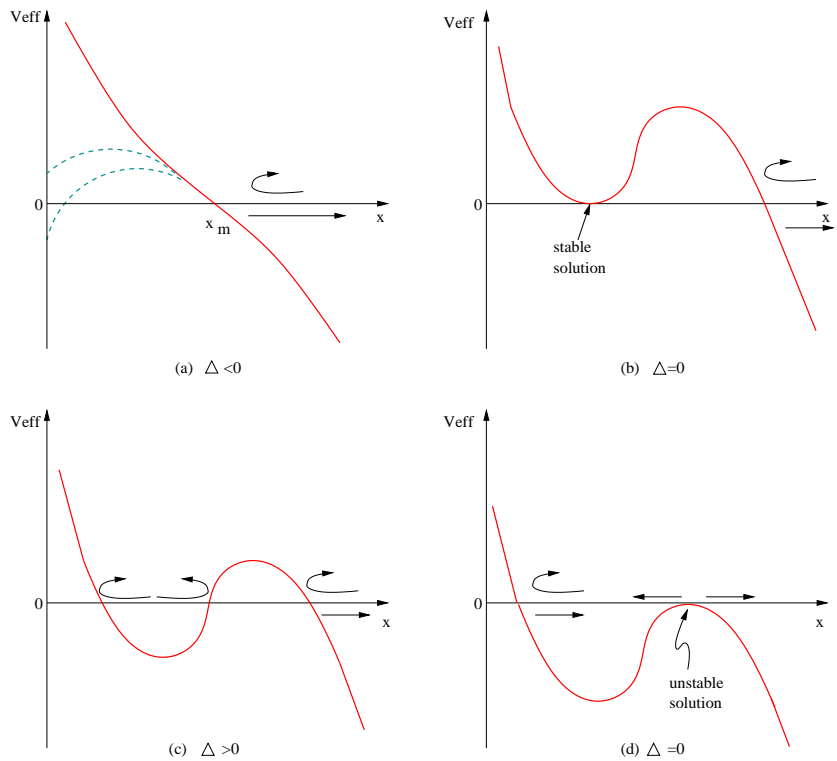
\fig{11cm}{10cm}{RSa8}{\small Plots of the effective potential for
$a_{8} < 0$ as a function of $x$. There are four cases: (a) a single
root, (b) a degenerate small root and a large root,  (c) three
distinct roots, and (d) a degenerate large root and a small root.
The arrows denote the possible trajectories of a shell with zero
energy. The dotted lines denote the non-rotating cases $(a_2=0)$,
which exist only for $\Delta <0$ defined in Eq. (\ref{eq:Delta}),
and there are two types, depending on the value of $a_4$. Note that
there is a stable equilibrium solution located at the local minimum
in (b), and an unstable equilibrium solution located at the local
maximum in (d).
 }{fig:efpa8}

For vanishing $a_8$, i.e., the same interior and exterior
geometries, the numerator of the effective potential reduces to a
quadratic polynomial in $x^2$. The coefficient of the highest order
term $a_6$ can have both positive and negative values, depending on
the values of parameters, and the shapes of the effective potential
are depicted in Fig. \ref{fig:effp}.

The crucial point concerning the different effective potentials
depicted in Figs. \ref{fig:efpa8} and \ref{fig:effp} is the
centrifugal barrier that appears around $x=0$. It is this barrier
that prevents the shell from contracting toward zero size, in turn
preventing the formation of a curvature singularity at $x=0$ unlike
the non-rotating case \cite{mo}.

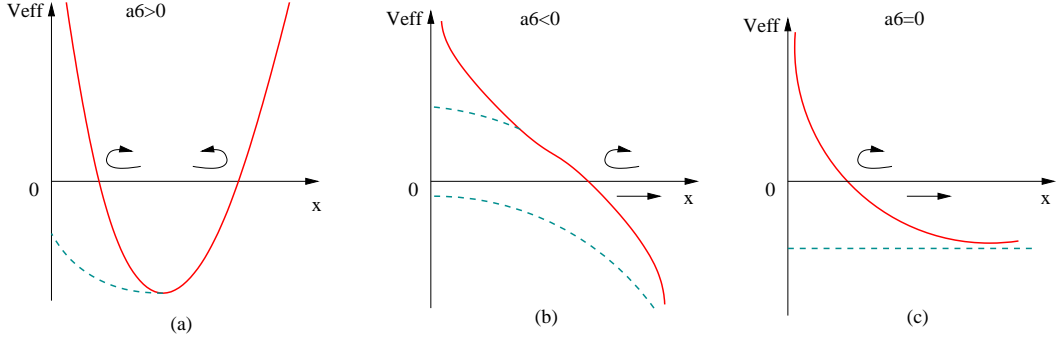
\fig{14cm}{4.5cm}{RSEP}{\small Plots of the effective potential  as
a function of $x$ for $a_{8}=0$. The arrows denote the possible
trajectories of a shell.  The dotted  lines denote the non-rotating
case $(a_2=0)$ and there can be two types, depending on the value of
$a_4$. }{fig:effp}

We now turn to a discussion of exact solutions.

\subsection{Exact Solutions I:  Different Interior/Exterior Geometries
$(a_8\neq0)$.}

We first consider the case of interior dS and exterior AdS spaces by
setting $\alpha_{o/i}^{+}=+1$, $\alpha_{o}^{-}=-\alpha_{i}^{-}=+1$,
$M_{+}=-(k^{+}_o +k_{i}^{+})$, $M_{-}=k^{-}_{o} -k_{i}^{-}$, $J=2
\ell \sqrt{ k^{\pm}_o k^{\pm}_i}$, then we get
\begin{eqnarray}
\label{a:kds/btz}
&&a_{8}=-4,\nonumber \\
&& a_6=4(M_+-M_-), \nonumber \\
&&a_{4}=-(\mu_{0}^2+M_+ + M_-)^2 + 4 M_+ M_-,\nonumber \\
&&a_{2}=J^2_{\pm} \mu_{0}^2/\ell^2.
\end{eqnarray}
We can also reverse the above case by setting $\alpha_{o/i}^{-}=+1$,
$\alpha_{o}^{+}=-\alpha_{i}^{+}=+1$, $M_{-}=-(k^{-}_o +k_{i}^{-})$,
$M_{+}=k^{+}_{o} -k_{i}^{+}$, $J=2 \ell \sqrt{ k^{\pm}_o
k^{\pm}_i}$, and easily find that the
coefficients are obtained by switching `$+$'
$\leftrightarrow$
`$-$', as expected due to the corresponding symmetry in the
configuration.

Next consider an interior dS and exterior flat space. This
corresponds to setting $\alpha_{o}^{+}=+1,~ \alpha_{i}^{+}=0$,
$\alpha_{o}^{-}=-\alpha_{i}^{-}=+1$,  $m_+=2 (1-\sqrt{k^+_i} ),~
j_+=2 \ell  \sqrt{k^+_o}$, $M_{-}=k^{-}_o -k_{i}^{-}$, $J_-=2 \ell
\sqrt{ k^{-}_o k^{-}_i}$, which reduces the coefficients to
\begin{eqnarray}
a_8&=& -1, \nonumber \\
a_{6}&=&-\left[\frac{m_{0}^{2}}{2} +\frac{1}{2 } (m_+ -2 )^2 +2M_-\right],\nonumber\\
a_4&=& -\left[\frac{m_{0}^2}{4}-\frac{1}{4 } (m_+ -2 )^2 +M_-
\right]^2 - (m_+ -2 )^2 M_-, \nonumber \\
a_2&=&  \frac{1}{16 \ell^2 } (m_+- 2 )^2 j_+^2 m_{0}^2=\frac{J_-^2
m_{0}^2}{4 \ell^2}.
\end{eqnarray}
In the last line, we have used the junction condition $k^+_o
k^+_i=k^-_o k^-_i$ and this shows that it is not the angular
momentum $  j_+ $ itself but the combination  $(2- m_+)  j_+ /2$  (with $m_+$
the mass parameter of the exterior locally flat space) that is continuous across the shell and matches with the angular momentum $ J_-$ of the interior dS
space. Note that the case of interior AdS and exterior flat spaces
can be similarly obtained by changing $a_6 \rightarrow -a_6$ in the
above formula with $M_{-}=-(k^{-}_o +k_{i}^{-})$. Reversing the
interior and exterior spaces can be obtained by switching `$+$'
$\leftrightarrow$ `$-$'.

Eq. (\ref{eq:eqnmot}) is not a particularly convenient form for
obtaining an exact solution. Rather, by introducing $X\equiv x^2$,
we find that (\ref{eq:eqnmot}) can be rewritten as
\begin{equation}
\dot{X}^2 + {\hat V}_{{\rm eff}}(X) = 0,
\end{equation}
where
\begin{equation}
\hat{V}_{{\rm eff}}(X) \equiv 4X V_{{\rm eff}}(x)=\frac{4}{m_{0}^2}
(a_8 X^3 + a_6 X^2 + a_4 X + {a_2}).
\end{equation}
Since the effective potential can be rewritten in the form $(a>0)$
\begin{equation}
\hat{V}_{{\rm eff}}(X) = {-} \frac{4}{m_{0}^2} a(X-b)(X-c)(X-d),
\end{equation}
where $a_{8}=-a$, $a_{6}= a(b+c+d)$, $a_{4}=-a(bd+bc+cd)$, and
${a_{2}=abcd}$, one finds that the differential equation has an
exact solution in terms of the Jacobi elliptic
function\footnote{The properties of the Jacobi elliptic functions are
presented in Refs. \cite{mo,as}.}
\begin{eqnarray}
t-{\mathcal C} &=& -\frac{m_{0}}{\sqrt{a(d-b)}}{\rm
JacobiSN^{-1}}\left[\sqrt{\frac{x^2-b}{c-b}},\sqrt{\frac{c-b}{d-b}}\right]\nonumber\\
&=&\frac{m_{0}}{\sqrt{a(d-b)}}{\rm
EllipticF}\left[\sqrt{\frac{x^2-b}{c-b}},\sqrt{\frac{c-b}{d-b}}\right],
\end{eqnarray}
which can be rewritten as
\begin{equation}
\label{eq:exsolution} x(t)=\left[b +(c-b){\rm
JacobiSN}^2\left\{\frac{\sqrt{a(d-b)}}{m_{0}}(t-{\mathcal C}),
\sqrt{\frac{c-b}{d-b}}\right\}\right]^{1/2}.
\end{equation}
The integration constant ${\mathcal C}$ is
\begin{eqnarray}
{\mathcal C} &=& \frac{m_{0}}{\sqrt{a(d-b)}}{\rm
JacobiSN^{-1}}\left[\sqrt{\frac{x_0^2-b}{c-b}},\sqrt{\frac{c-b}{d-b}}\right]
\nonumber \\
 &=&\frac{m_{0}}{\sqrt{a(d-b)}}{\rm
EllipticF}\left[\sqrt{\frac{x_0^2-b}{c-b}},\sqrt{\frac{c-b}{d-b}}\right]
\end{eqnarray}
determined by setting
$x=x_0$ at $t=0$.

The three roots $b$, $c$,
$d$ can be rewritten as
\begin{eqnarray}
 r_0 = s_+ +s_- -\frac{a_6}{3 a_8},~~  r_\pm =  -\frac{1}{2} (s_+ +s_-)
 -\frac{a_6}{3 a_8} \pm i \frac{\sqrt{3}}{2} (s_+ -s_-),
\end{eqnarray}
where
$s_{\pm} =\sqrt[3]{r \pm i \sqrt{q^3-r^2}}$
with
\begin{equation}
r=\frac{9 a_8 a_6 a_4-27 a_8^2 a_2 -2 a_6^3}{54 a^3_8},
~~q=\frac{-3 a_8 a_4 +a_6^2}{9 a_8^2}
\end{equation}
and  here the choice of identification of $\{ r_0, r_\pm\}$ with $\{
b,c,d \}$ is arbitrary. For $a_8 \neq 0$, the number of ``real''
roots depends on the discriminant
\begin{eqnarray}
\label{eq:Delta}
\Delta &\equiv &108 a_8^4 (q^3 -r^2) \nonumber \\
&=& -27a_2^2 a_8^2 + 2 a_4(9 a_6 a_2 -2 a_4^2)a_8 + a_6^2(a_4^2-4a_6 a_2).
\end{eqnarray}
If $\Delta >0$, there are three distinct real roots and if
$\Delta=0$, (at least) two roots coincide, while if $\Delta
<0$, there is only one real root with a pair of complex conjugate
roots (Fig. 1). Plugging Eq. (\ref{a:kds/btz}) into Eq.
(\ref{eq:Delta}) yields
\begin{eqnarray}
\Delta &=&-27 \mu_{0}^{12} -162 \mu_{0}^{10} z -54 \mu_{0}^8
(y^2+6z^2)-54\mu_{0}^6 (4(z y^2+z^3) -9J^2 y) \nonumber \\
&&-27 \mu_{0}^4 (y^4 +8 z^2 y^2+27 J^4 -36 J^2 z y)-54 \mu_{0}^2 (z
y^4- J^2 y^3)\label{Deltaeqn}
\end{eqnarray}
with   $\mu_{0} \equiv m_0/2$, $y \equiv M_{+}
-M_{-}$, and $z \equiv M_{+} + M_{-}$.

Thus we conclude that collapse scenarios with different interior and
exterior geometries (see Fig. \ref{fig:efpa8}) are described by the
exact solution of Eq. (\ref{eq:exsolution}), with parameters
appropriately chosen. Note that  for the non-rotating case ($J=0$
and $z>0$) $\Delta <0$ always (Fig. \ref{fig:efpa8} (a)).  The
crucial difference between the rotating case and the non-rotating
case \cite{mo} is the centrifugal  barrier around the origin that
prevents the shell from collapsing to zero size, forbidding the
formation of a curvature singularity for the former. This is
manifest in Eq. (\ref{Deltaeqn}), where we can see that the effect
of rotation gives {\it positive} contributions for $y>0$, and so can
render $\Delta$ non-negative (Fig. \ref{fig:efpa8} (b), (c), and
(d)).

\subsection{Exact Solutions II:  Same Interior/Exterior Geometries $(a_8=0)$.}

For the same interior and exterior geometries we have $a_8=0$. The
solutions become simpler in that they can be expressed by
trigonometric or exponential functions. To see this, we first
consider   AdS spaces in both regions by setting
$\alpha_{o/i}^{\pm}=+1$. Then we get
\begin{eqnarray}
&&a_{8}=0, ~a_6=m_{0}^2, \nonumber \\
&&a_{4}=-[m_{0}^2/4-(k_{o}^{+}+k_{i}^{+}+k_{o}^{-}+k_{i}^{-})]^2 +
4(k_{o}^{+}+k_{i}^{+})(k_{o}^{-}+k_{i}^{-}),\nonumber \\
&&a_{2}=k_{o}^{+}k_{i}^{+}m_{0}^2
\end{eqnarray}
with the condition $k_{o}^{+}k_{i}^{+}=k_{o}^{-}k_{i}^{-}$.
Alternatively, if we define the mass and angular momentum parameters
$M_{\pm}\equiv -(k_{o}^{\pm}+k_{i}^{\pm})$ and $J \equiv 2 \ell
\sqrt{k_{o}^{\pm}k_{i}^{\pm}}$, then the black hole and the AdS
point mass spacetimes can be described by $M>0$ and $M<0$,
respectively, and one finds
\begin{equation}
V_{{\rm eff}}(x) = \frac{m_{0}^2 x^4 - \tilde{a}_4 x^2 + m_{0}^2
J^2/4\ell^2}{m_{0}^2 x^2},
\end{equation}
where $\tilde{a}_4\equiv -a_4$ is positive definite when at least
one of the interior/exterior geometries is a black hole geometry.
For example for BTZ black holes in both regions $(M_{\pm}>0)$, we
have $\tilde{a}_4= m_{0}^4/16 +m_{0}^2 (M_+ +M_-)/2 +(M_+ - M_-)^2
>0$. Likewise for interior AdS point mass and exterior BTZ black hole case
$(M_+ M_-<0)$ we have $\tilde{a}_4=(m_{0}^2/4 + M_{+}+M_{-})^2
-4M_{+}M_{-}>0$. However for AdS point masses in both regions,
$\tilde{a}_4$ cannot be positive always.

The shape of the effective potential is given in Fig. \ref{fig:effp}
(a) since $a_{6} = 4\mu_{0}^2 >0$. So there are two positive roots
at $x=x_{{\rm min/max}}$, where
\begin{equation}
x_{{\rm min}} = \frac{\sqrt{\tilde{a}_4 - \sqrt{\tilde{a}_4^2 -
m_{0}^4 J^2/\ell^2}}}{\sqrt{2}m_{0}},~~x_{{\rm max}} =
\frac{\sqrt{\tilde{a}_4 + \sqrt{\tilde{a}_4^2 - m_{0}^4
J^2/\ell^2}}}{\sqrt{2}m_{0}}.
\end{equation}
and the shell moves between them as illustrated in Fig.
\ref{fig:effp} (a). An exact solution can be found by a
straightforward computation:
\begin{equation}
t-{\mathcal C} = -\frac{1}{2} {\arctan}\left[ \frac{\tilde{a}_4-
2m_{0}^2 x^2}{2\sqrt{- m_{0}^4 x^4 + \tilde{a}_4 m_{0}^2 x^2 -
m_{0}^2 J^2/\ell^2}}\right].
\end{equation}
Alternatively we can write
\begin{equation}
\label{eq:solx} x(t) = \frac{1}{\sqrt{2} m_{0}} \left[ \tilde{a}_4 +
\sqrt{\tilde{a}_4^2-m_{0}^4 J_{+}^2} ~|\sin 2(t-{\mathcal
C})|\right]^{1/2}
\end{equation}
or
\begin{equation}
\label{eq:solx_2} x(t) = \frac{1}{\sqrt{2}} \left[(x^2_{{\rm
max}}+x^2_{{\rm min}})+(x^2_{{\rm max}}-x^2_{{\rm min}}) ~|\sin
2(t-{\mathcal C})|\right]^{1/2},
\end{equation}
where the integration constant ${\mathcal C}$ is determined by an
initial condition on the position of the shell $x_{0}\equiv x(t=0)$,
\begin{equation}
{\mathcal C} = \frac{1}{2} {\arctan}\left[ \frac{\tilde{a}_4-
2m_{0}^2 x_0^2}{2\sqrt{-m_{0}^4 x_0^4 + \tilde{a}_4 m_{0}^2 x_0^2 -
 m_{0}^2 J^2/\ell^2}}\right].
\end{equation}
It is clear that a (real) solution exists only for $\tilde{a}_4>0$:
For AdS point masses in both regions, i.e., $M_{\pm} <0$, we need
particular initial configurations satisfying $m_{0}^2/4 > 2
\sqrt{M_{+} M_{-}} -(M_{+} + M_{-})$ or $m_{0}^2/4 < -2
\sqrt{M_{+}M_{-}} -(M_{+} + M_{-})$ in order to have real solutions.

Note that the minimum bound at $x=x_{\rm min}$ arises due to the
angular momentum parameter $J$, which implies that the shell will
not shrink to zero size. The intrinsic Ricci scalar on the shell
computed from the induced metric (\ref{eq:indmet}),
\begin{equation}
{\cal R}_{\mu}^{\mu}[\Sigma] =
\frac{2}{x(t)}\frac{d^2x(t)}{dt^2} = - \frac{1}{x}\frac{d V_{{\rm
eff}}}{dx}, \label{eq:ricci}
\end{equation}
has no singularity since the shell cannot
reach the origin at $x=0$ due to the angular momentum barrier. This
implies that rotational effects prevent the curvature singularity at
$x=0$ from being formed.

Formation of a BTZ black hole will take place if certain
initial conditions hold. First the inequalities $x_{\rm
max} \geq  x_0 > x_H^+ >x_{\rm min}$  must hold, so that the shell's initial
location is outside of the putative  horizon $x_H^+$ of the exterior
spacetime, which in turn must be located between $x_{\rm min}$ and
$x_{\rm max}$.  A BTZ black hole can form for the exterior observer
before the shell moving inward either collapses {\it onto} the
interior point mass, collapses {\it into} a point mass if the
interior is pure AdS, or  is absorbed into the outer horizon of an
interior BTZ black hole\footnote{ The formation
of a black hole would also occur for interior flat or AdS point
mass spacetime provided appropriate conditions are met.}.

To see this,
consider Eq. (\ref{Junction1}), which can be written as
\begin{equation}
\label{Junction1a} \frac{m_{0}}{2} =\sqrt{
{\dot{x}^2}+x^2+\frac{J^2}{4 \ell^2 x^2} -M_{-} } - \sqrt{
{\dot{x}^2}+x^2+\frac{J^2}{4 \ell x^2} -M_{+}}
\end{equation}
and note that everything in the square-root terms is the same except
for $M_{+}$ and $M_{-}$.

If there is a black hole in the interior spacetime, i.e., $M_->0$,
then a black hole in the exterior spacetime, i.e., $M_+>0$, will
always form for a positive $m_0$, for any initial point $x_0$ : The
shell moving inward forms the (outer) black hole horizon before
being absorbed by the outer horizon of the interior black hole,
i.e., $x^-_H < x^+_H$ which is equivalent to $M_-<M_+$.

If the interior is pure AdS or has a point
mass, i.e., $M_-<0$, then a black hole in the exterior spacetime ($M_+>0$)
forms if
\begin{equation}
\bar{m}_0 < m_0
\end{equation}
with $\bar{m}_0/2= \sqrt{{{\dot{x}}_{0}^2}+{{\cal N}^2_{-}(x_0)}} -
\sqrt{{{\dot{x}}_0^2}+{{\cal N}^2_{+}(x_0)} +M_+}$, for any initial
point $x_0$.

For ${m_0}<\bar{m}_0$ then $M_+ < 0$ and the
shell collapses into (or onto) a point mass, which is similar to the
condition for dust {\it cloud} collapse in the non-rotating case
\cite{mr}. Furthermore, for $0<m_0 <\bar{m}_0$ we have
$0>M_{+}>M_{-}$; the deficit angle, which is defined by $2 \pi
(1-\sqrt{-M_{\pm})}$ \cite{Carl:98}, of the point mass outside can
not be smaller than that of the point mass inside. Note that the
physics governing the endstate of collapse is basically the same
regardless of the values of $\dot{x}_0$ and $J$.

On the other hand, by squaring (\ref{Junction1}) one find that the
shell's gravitational mass $M_+$, which becomes the black hole mass
after its formation, is given by
\begin{eqnarray}
 M_+=m_0 \sqrt{\dot{x}^2+{\cal
N}_-^2} -2 G m_{0}^2  +M_-,
\end{eqnarray}
(reinstating Newton's constant $G$) where the first and second
terms correspond to the shell's relativistic kinetic energy and
binding energy, respectively \cite{poisson}. Note that the binding
energy is negative only for  {\it positive} $G$,
as in the conventional higher dimensional black holes
\cite{Myer:86}\footnote{Constant binding energy
is peculiar to three dimensions; in general there is distance dependence. In four
dimensions, for example, the gravitational mass
of a spherically symmetric shell, with a flat interior spacetime, is
given by $M_+=m_0 \sqrt{\dot{x}^2+1} -G_4 m_{0}^2/2R$ with $G_4$ the
four-dimensional Newton's constant. }; this is unique to AdS
spacetimes and this could provide a physical explanation of why the
black hole solution can exist only in this case, but not in dS or
flat spacetimes.

Next we consider dS spaces in both regions by setting
$\alpha_{o}^{\pm}=+1$ and $\alpha_{i}^{\pm}=-1$. Then we get
\begin{eqnarray}
&&a_{8}=0, ~a_6=-m_{0}^2, \nonumber \\
&&a_{4}=-[m_{0}^2/4+(k_{o}^{+}-k_{i}^{+}+k_{o}^{-}-k_{i}^{-})]^2 +
4(k_{o}^{+}-k_{i}^{+})(k_{o}^{-}-k_{i}^{-}),\nonumber \\
&&a_{2}=k_{o}^{+}k_{i}^{+}m_{0}^2
\end{eqnarray}
with the condition $k_{o}^{+}k_{i}^{+}=k_{o}^{-}k_{i}^{-}$. The
effective potential is  ($\tilde{a}_4\equiv -a_4>0$)
\begin{equation}
\label{V:eff:dsdS}
 V_{{\rm eff}}(x)=\frac{-m_{0}^2 x^4 -\tilde{a}_4
x^2 + m_{0}^2 J^2/4\ell^2}{m_{0}^2 x^2}.
\end{equation}
In this case, the shape of the effective potential is Fig. 2(b)
since $a_6=-m_{0}^2 <0$ with a single root at $x=x_{{\rm min}}$,
where
\begin{eqnarray}
\label{x:min:dsds}
 x_{{\rm min}}= \frac{\sqrt{-\tilde{a}_4 +
\sqrt{\tilde{a}_4^2 + m_{0}^4J^2/\ell^2}}}{\sqrt{2}m_{0}}
\end{eqnarray}
provided $J=2 \ell \sqrt{k_{o}^{\pm}k_{i}^{\pm}} \neq 0$; for
vanishing angular momentum $(J=0)$, i.e., non-rotating dS spaces,
there is no $x_{{\rm min}}$ (the lower dotted line in Fig. 2(b)).
The solution for the shell's edge is
\begin{equation}
x(t) = \frac{1}{2 m_{0}}\sqrt{\frac{1}{\ell^2} m_{0}^2 J^2
e^{2(t-{\mathcal C})} + \left[m_{0} e^{-(t-{\mathcal C})} -
\frac{\tilde{a}_4}{m_{0}}e^{(t-{\mathcal C})}\right]^2},
\end{equation}
where the integration constant ${\mathcal C}$ is
\begin{equation}
{\mathcal C} = -\frac{1}{2}\ln\left[ \frac{8 m_{0}^2 x_0^2 + 4
\tilde{a}_4 + 8 m_{0}\sqrt{m_{0}^2 x_0^4 + \tilde{a}_4 x_0^2 -
m_{0}^2 J^2/4\ell^2}}{m_{0}^2}\right].
\end{equation}
As with black hole spacetimes, when some appropriate conditions are
imposed, a (cosmological) event horizon $x_{C}^-$ can form from the
perspective of the interior observer. If the cosmological horizon
 $x_{C}^-$ of the interior spacetime is located
larger than $x_{{\rm min}}$ and if the initial location of the dust
shell is in between $x_{{\rm min}}$ and $x_C^-$, the expanding (or
collapsing and later expanding) shell will form a cosmological
horizon for the interior (KdS$_3$) observer.

To see this explicitly, we note that, as in the black hole case, the
cosmological horizons are located always larger than $x_{\rm
min}$\footnote{This can be also understood from (\ref{eq:eqx}) which
is always satisfied when the shell's location coincides with one of
the horizons, i.e., ${\cal N}_+=0$ or ${\cal N}_-=0$. }, i.e.,
$x_{\rm min} \leq x_C^{\pm}=\sqrt{k_o^{\pm}}$, from ${\tilde{a}_4
+m_{0}^2 M_{\pm}}=(m_{0}/2)^4 +m_{0}^2 (M_{\mp}+3 M_{\pm})/2+(M_+
-M_-)^2 \geq 0 $ for any non-zero $\mu_{0}$. Regardless of the sign
of  $(M_+ +3 M_-)$ this condition is trivially satisfied since there
are no real values of $\mu_{0}$ satisfying ${\tilde{a}_4 +m_{0}^2
M_{\pm}} < 0$.

Consider again  Eq. (\ref{Junction1}), which we write as
\begin{equation}
\label{Junction1a} \frac{m_{0}}{2} =\sqrt{
{\dot{x}^2}+M_{-}+\frac{J^2}{4 \ell^2 x^2}-x^2  } - \sqrt{
{\dot{x}^2}+M_{+}+\frac{J^2}{4 \ell x^2} -x^2}.
\end{equation}
Note that the contributions of the mass terms $M_{\pm}$ are opposite
to those of the black hole spacetime due to our definition of mass
\cite{Park:98,Bala:01}. Then  the expanding (or collapsing and later
expanding) shell will form a cosmological horizon for the interior
(KdS$_3$) observer for a {\it negative} $m_0$: If $m_0>0$, then
$x^-_C > x^+_C$, which is equivalent to $M_+>M_-$, and the shell
will be absorbed by the cosmological horizon of the exterior
spacetime, before forming the cosmological horizon in the interior
spacetime\footnote{ In the non-rotating case this
scenario has been previously noted in a higher-dimensional context
\cite{MyHub}.}.

Finally we consider  flat spaces in both regions by setting
$\alpha_{i}^{\pm}=0$ and $\alpha_{o}^{\pm}=+1$. We have
\begin{eqnarray}
&&a_8=a_6=0, \nonumber \\
&&a_{4}=-[m_{0}^2/4-(k_{i}^{+}+k_{i}^{-})]^2 +
4k_{i}^{+}k_{i}^{-},\nonumber \\
&&a_{2}=k_{o}^{+}k_{i}^{+}m_{0}^2
\end{eqnarray}
with the condition $k_{o}^{+}k_{i}^{+}=k_{o}^{-}k_{i}^{-}$. If we
define the mass and angular momentum
 parameters $m^{\pm}$ and $j^{\pm}$ of the point masses
to be  $k^{\pm}_{i}=(m_{\pm}-2 )^2/4 $, $k^{\pm}_o =j^2_{\pm}/4
\ell^2$,
 then one finds the effective potential becomes
\begin{equation}
V_{{\rm eff}}(x)=\frac{-\tilde{a}_4 x^2 + ({16 \ell^2})^{-1}
(m_{\pm}-2)^2j_{\pm}^2 m_{0}^2}{m_{0}^2 x^2},
\end{equation}
where $\tilde{a}_4\equiv -a_4=[m_{0}^2/4 -(k^+_i + k^-_i)]^2 -4
k^+_i k^-_i$. Its shape is shown in Fig. \ref{fig:effp}(c), since
$a_6=0$ with a single root at $x=x_{{\rm min}}$, where
\begin{eqnarray}
x_{{\rm min}}=\frac{|(m_{\pm}-2 ) j_{\pm} | m_{0}}{4 \ell
\sqrt{\tilde{a}_4}}.
\end{eqnarray}
Here we note that the angular momenta $j_{\pm}$ are not separately
continuous across the shell, but only  the combinations $(m_{\pm}-2
) j_{\pm}$ are. The exact solution is found to be
\begin{equation}
\label{eq:solflat} x(t) = \frac{m_{0}|(m_{\pm}-2)j_{\pm}|} {4 \ell
\sqrt{\tilde{a}_4}}\sqrt{1+\frac{16 \tilde{a}_4^2
\ell^2}{(m_{\pm}-2)^2 {j_{\pm}^2m_{0}^4}} (t-{\mathcal C})^2},
\end{equation}
where the integration constant ${\mathcal C}$ is
\begin{equation}
{\mathcal C}=\frac{m_{0}}{2 \tilde{a}_4} \sqrt{4\tilde{a}_4 x_0^2
-(m_{\pm}-2)^2  j_{\pm}^2m_{0}^2/4 \ell}.
\end{equation}

To summarize this section, we find that rotating dust shells can
collapse,  but only down to a minimal size $x_{{\rm min}}$ (after
which they expand) due to their angular momentum. No curvature
singularity is formed during the gravitational collapse process,
unlike the non-rotating case \cite{mo}. The alternative collapse
endpoint is a BTZ black hole or a KdS$_3$ spacetime.

\section{Shells with Pressure}\label{sec:pressure}

In this section, we shall consider shells with pressure determined
by somewhat generalized equations of state.  The effect of pressure
is to produce a varying shell energy $2 \pi \ell \rho x$ from Eq.
(\ref{eq:eqn2}), i.e., deviations from the inverse radius dependence
of $\rho = m_0 /2 \pi \ell x$. Specifically we consider shells with
the polytropic-type equation of state
\begin{equation}
p=\frac{\omega \rho}{\pi} \left(\frac{2 \pi \ell
\rho}{m_{0}}\right)^{1/n} ,\label{eq:poly}
\end{equation}
that encompasses many sorts of known fluids by choosing a specific
equation of state parameter $\omega$ and polytropic index $n$. For
instance, we have constant energy density ($n=0$), non-relativistic
degenerate fermions ($n=1$), non-relativistic matter or radiation
pressure ($n=2$), and linear (perfect)  fluid ($n\rightarrow
\infty$), respectively \cite{cf}. Moreover, the equation of state in
Eq. (\ref{eq:poly}) can describe a Chaplygin gas by choosing $n,
\omega <0$.

Plugging the equation of state (\ref{eq:poly}) into Eq.
(\ref{eq:eqn2}) yields
\begin{equation}
\label{rho:finite_n} \rho(x) =\frac{m_0 }{2 \pi \ell}
\left(-\omega+\mu x^{1/n}\right)^{-n},
\end{equation}
where $\mu$ is an integration constant. Similarly, for the
 linear fluid ($n \rightarrow \infty$), we have
\begin{equation}
\rho(x) = \frac{m_0 }{2 \pi \ell} x^{-(\omega+1)}.
\end{equation}
From these we obtain the equations
\begin{equation}
\sqrt{\dot{x}^2+{\mathcal N}_{+}^2}-\sqrt{\dot{x}^2+{\mathcal
N}_{-}^2} + \frac{( m_{0}x/2 )}{(-\omega+(1+\omega)x^{1/n})^{n}} =
0,
\end{equation}
for finite $n$, and
\begin{equation}
\sqrt{\dot{x}^2+{\mathcal N}_{+}^2}-\sqrt{\dot{x}^2+{\mathcal
N}_{-}^2} + \frac{(m_{0}/2)}{x^{\omega}} = 0,
\end{equation}
for the linear  fluid, where $\mu=1+\omega$ so that $\rho={m_0 }/{2
\pi \ell}$ when $x=1$.

For finite $n$ and the linear  fluid,  the
equations of motion can be expressed in the alternate form
\begin{equation}
\dot{x}^2 + V_{{\rm eff}}(x) = 0,
\end{equation}
where the effective potential is
\begin{equation}
\label{V_eff:p}
 V_{{\rm eff}}(x) = \frac{1}{2}({\mathcal
N}_{+}^2+{\mathcal N}_{-}^2)-\frac{\pi^2 \ell^2}{4}\rho^2 x^2
-\frac{1}{4 \pi^2 \ell^2 \rho^2 x^2}({\mathcal N}_{+}^2-{\mathcal
N}_{-}^2)^2
\end{equation}
with the junction condition $k_{o}^{+}k_{i}^{+}=k_{o}^{-}k_{i}^{-}$.
Note that the only change to the effective potential compared to
pressureless dust shells in Eq. (\ref{veff1}) is the replacement of
$m_{0} \rightarrow 2 \pi \ell \rho x$ in Eq. (\ref{veff1}).

Up to now, the preceding results are  valid for any $n$. But the
behaviour of the effective potential differs for $n>0$ and $n<0$,
which need separate consideration.

\subsection{The Ordinary Polytropic Shells: $n, \omega>0$}\label{sec:poly_ord}

For ordinary fluids with finite and positive definite $n$ and
$\omega$, the effective potential and the intrinsic Ricci scalar
also {\it negatively} diverge at $x=x_{\omega} \equiv
\omega^{n}/(1+\omega)^n$, where the density $\rho$ also diverges. So
the shell will not collapse to a point but rather to a ring of
finite size $x=x_{\omega}$ in a finite proper time. The shape of the
effective potential is depicted in Figs. \ref{fig:numpics} and
\ref{fig:numpics2}.

On the other hand, for the linear  fluid ($n \rightarrow \infty$)
there is no ring singularity since $\rho$ is finite for any finite
$x$. Rather, from the effective potential near $x=0$,
\begin{eqnarray}
\label{eq:VnearX0} V_{{\rm eff}} &\approx& \frac{J^2}{4 \ell^2 x^2}
-\frac{m_{0}^2}{16 x^{2 \omega}} + \frac{1}{2} (\alpha_{i}^{+}
k_{o}^{+}+\alpha_{o}^{+} k_{i}^{+}+\alpha_{i}^{-}
k_{o}^{-}+\alpha_{o}^{-}
k_{i}^{-})\nonumber \\
&&-\frac{x^{2 \omega}}{ m_{0}^2} (\alpha_{i}^{+}
k_{o}^{+}+\alpha_{o}^{+} k_{i}^{+}-\alpha_{i}^{-}
k_{o}^{-}-\alpha_{o}^{-} k_{i}^{-})^2,
\end{eqnarray}
due to  $\rho(x) =({m_0 }/{2 \pi \ell})x^{-(\omega+1)}$, a shell
with $\omega >1$ will collapse to zero size\footnote{ The case
$\omega<-1$ shows similar behaviour for the effective potential and
$\omega=-1$ is a marginal case. We shall not consider these
possibilities, i.e., $\omega<0$ with the usual values of $n>0$,
since their physical relevance is not quite clear. }, {\it
regardless of} the initial values of $m_{0}$ and $k^{\pm}_o
k^{\pm}_i \equiv J^2/4\ell^2$, in a finite proper time where the
effective potential and the energy density (or pressure) diverge
(Figs. \ref{fig:numpics6} and \ref{fig:numpics7}).

Provided the exterior geometry has no black hole (event) horizon
such as the KdS$_3$ space or the AdS/flat spaces with point  masses,
there is nothing to prevent the collapsing shell from developing a
curvature singularity. The resultant singularity is naked and we
have a violation of the cosmic censorship for ``generic'' initial
data. These are qualitatively the same behaviours as those of
non-rotating shell collapse \cite{mo}, which implies that their
angular momentum is not large enough to overcome forming the
curvature singularity for most fluids whose equations of state are
of the form (\ref{eq:poly}) . The centrifugal barrier that prevents
collapse in the pressureless dust case still occurs at $x=0$, but is
always dominated by the negatively divergent effective potential at
the finite value of $x$.

However, for the linear  fluid with $\omega <1$, angular momentum
effects again dominate, ensuring that cosmic censorship is upheld
regardless of the relative values of $m_{0}$ and $J$ (Figs.
\ref{fig:numpics8} and \ref{fig:numpics9}). Note that this is the
case where there is a similarity with the pressureless dust shells
in Sec. 3: Actually, Figs. \ref{fig:numpics8} and \ref{fig:numpics9}
(and Fig. \ref{fig:numpics5} as well) show all the possible cases in
Figs. \ref{fig:efpa8} and \ref{fig:effp}.

The case  $\omega=1$ is a marginal case that crucially depends on
the initial  data: One has an infinite well for $m_{0}^2>4J^2$ (Fig.
\ref{fig:numpics3}) but an infinite barrier for $m_{0}^2 <4J^2$
(Fig. \ref{fig:numpics5}); for  $m_{0}^2 =4J^2$, the point $x=0$ is
naked when $(\alpha_{i}^{+} k_{o}^{+}+\alpha_{o}^{+}
k_{i}^{+}+\alpha_{i}^{-} k_{o}^{-}+\alpha_{o}^{-} k_{i}^{-})<0$
(Fig. \ref{fig:numpics4}).

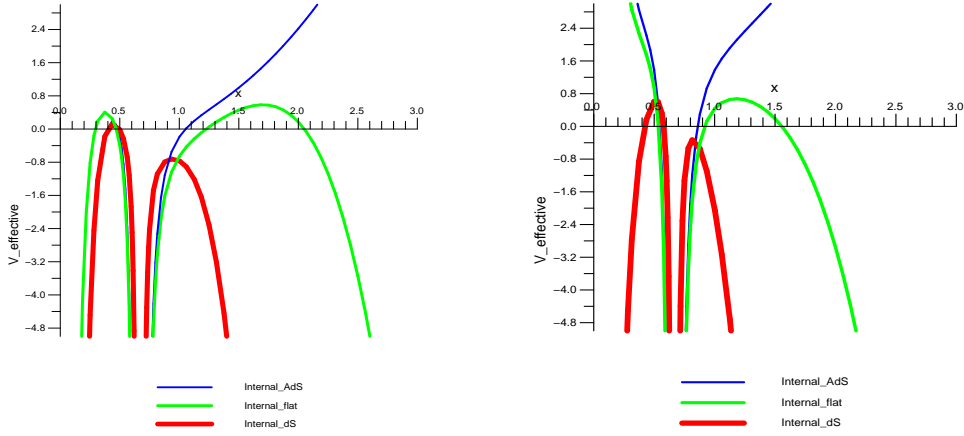
\figs{7cm}{6.3cm}{Eadsbh}{7cm}{6.3cm}{Eadspm}{\small Plots of the
effective potentials for the shell with a polytropic equation of
state and the exterior AdS black hole
($\alpha_{o/i}^{+}={+}1,k_{o}^{+}= -(x^+_H)^2=-1,k_i^{+}=-1/4$:
LHS)/AdS point mass ($\alpha_{o/i}^{+}=
+1,k_{o}^{+}=+1,k_i^{+}=+1/4$: RHS). We choose an interior AdS space
with a point mass ($\alpha_{i}^{-}=+1$), flat ($\alpha_{i}^{-}=0$),
and dS ($\alpha_{i}^{-}=-1$) by setting parameters
$\alpha_{o}^{-}=+1, \omega=+2,k_{o/i}^{-}=+1/2, {m_{0}=4},
n=+1$.}{fig:numpics}

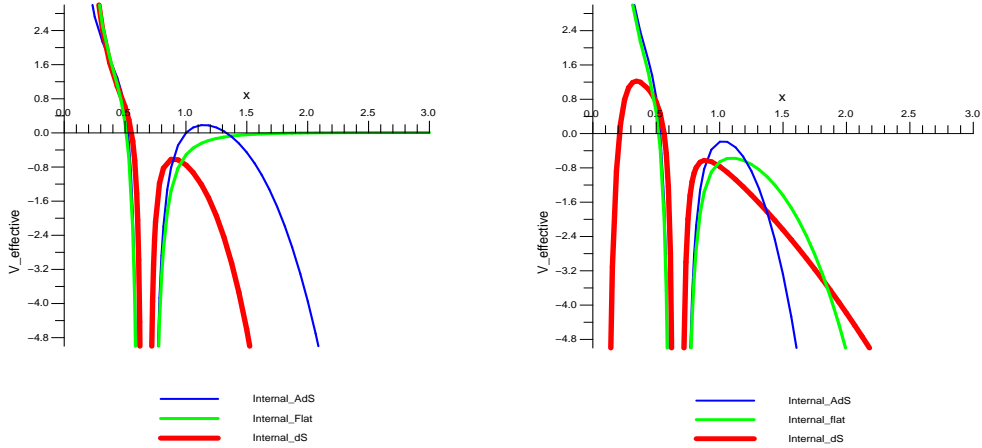
\figs{7cm}{6.5cm}{Eflat}{7cm}{6.5cm}{Eds}{\small Plots of the
effective potentials for the shell with a polytropic equation of
state and the exterior flat space ($\alpha^+_o=0,\alpha^+_i=+1,
k_{o}^{+}=+1,k^+_i=+1/4$: LHS)/dS space ($\alpha^+_{o}=+1,
\alpha^-_{i}=-1, k_{o}^{+}= (x^+_C)^2=+1,
k^+_i=+1/4$: RHS). We choose an interior AdS space with a point mass
($\alpha_{i}^{-}=+1$), flat ($\alpha_{i}^{-}=0$), and dS
($\alpha_{i}^{-}=-1$) by setting parameters $\alpha_{o}^{-}=+1,
\omega=+2, k_{o/i}^{-}=+1/2, m_{0}=2, n=+1$.}{fig:numpics2}


\figs{7cm}{7cm}{EadsVgpf2}{7cm}{7cm}{EadspmVgpf2}{\small Plots of
the effective potentials for the shell with a linear fluid and the
exterior AdS black hole ($k_{o}^{+}=
-(x^+_H)^2 =-1,k_i^{+}=-1/4$: LHS)/AdS point mass
($k_{o}^{+}=+1,k_i^{+}=+1/4$: RHS). We choose an interior AdS space
with a point mass ($\alpha_{i}^{-}=+1$), flat ($\alpha_{i}^{-}=0$),
and dS ($\alpha_{i}^{-}=-1$) by setting parameters
$\alpha_{o}^{-}=+1, \omega=+{2}, k_{o/i}^{-}=+1/2,
{m_{0}=4}$.}{fig:numpics6}

\figs{7cm}{7cm}{EflatVgpf2}{7cm}{7cm}{EdsVgpf2}{\small Plots of the
effective potentials for the shell with a linear fluid and the
exterior flat space ($\alpha^+_o=0,\alpha^+_i=+1,
k_{o}^{+}=+1,k^+_i=+1/4$: LHS)/dS space
($\alpha^+_{o/i}=+1,k_{o}^{+}=(x^+_C)^2 =+1,k^+_i=+1/4$: RHS). We
choose an interior AdS space with a point mass
($\alpha_{i}^{-}=+1$), flat ($\alpha_{i}^{-}=0$), and dS
($\alpha_{i}^{-}=-1$) by setting parameters,
$\alpha_{o}^{-}=+1,\omega=+2$, $k_{o/i}^{-}=+1/2$,
${m_{0}=4}$.}{fig:numpics7}

\figs{7cm}{7cm}{EadsbhVgpfhalf}{7cm}{7cm}{EadspmVgpfhalf}{\small
Plots of the effective potentials for the shell with a linear  fluid
and the exterior AdS black hole
($k_{o}^{+}=-(x^+_H)^2=-1,k_i^{+}=-1/4$: LHS)/AdS point mass
($k_{o}^{+}=+1,k_i^{+}=+1/4$: RHS). We choose an interior AdS space
with a point mass ($\alpha_{i}^{-}=+1$), flat ($\alpha_{i}^{-}=0$),
and dS ($\alpha_{i}^{-}=-1$) by setting parameters
$\alpha_{o}^{-}=+1, \omega=+1/2, k_{o/i}^{-}=+1/2, {m_{0}=4}$. The
event horizon  for the exterior observer is located at $x=1.0$ in
the left diagram. However a fluid collapsing through this surface
will not bounce from the interior barrier. }{fig:numpics8}

\figs{7cm}{7cm}{EflatVgpfhalf}{7cm}{7cm}{EdsVgpfhalf}{\small Plots
of the effective potentials for the shell with a linear  fluid and
the exterior flat space ($\alpha^+_o=0,\alpha^+_i=+1,
k_{o}^{+}=+1,k^+_i=+1/4$: LHS)/dS space
($\alpha^+_{o/i}=+1,k_{o}^{+}= (x^+_C)^2 =+1,k^+_i=+1/4$: RHS). We
choose an interior AdS space with a point mass
($\alpha_{i}^{-}=+1$), flat ($\alpha_{i}^{-}=0$), and dS
($\alpha_{i}^{-}=-1$) by setting parameters,
$\alpha_{o}^{-}=+2,\omega=+{1/2}$, $k_{o/i}^{-}=+1/2$,
${m_{0}=4}$.}{fig:numpics9}

\figs{7cm}{7cm}{EadsVgpf}{7cm}{7cm}{EadspmVgpf}{\small Plots of the
effective potentials for the shell with a linear fluid and the
exterior AdS black hole ($k_{o}^{+}=
-(x^+_H)^2 =-1,k_i^{+}=-1/4$: LHS)/AdS point mass
($k_{o}^{+}=+1,k_i^{+}=+1/4$: RHS). We choose an interior AdS space
with a point mass ($\alpha_{i}^{-}=+1$), flat ($\alpha_{i}^{-}=0$),
and dS ($\alpha_{i}^{-}=-1$) by setting parameters
$\alpha_{o}^{-}=+1, \omega=+1, k_{o/i}^{-}=+1/2,
{m_{0}=4}$.}{fig:numpics3}

\figs{7cm}{7cm}{Eflatpfbar}{7cm}{7cm}{Edspfbar}{\small Plots of the
effective potentials for the shell with a linear fluid and the
exterior flat space ($\alpha^+_o=0,\alpha^+_i=+1,
k_{o}^{+}=+1,k^+_i=+1/4$: LHS)/dS space
($\alpha^+_{o/i}=+1,k_{o}^{+}= (x^+_C)^2 =+1,k^+_i=+1/4$: RHS). We
choose an interior AdS space with a point mass
($\alpha_{i}^{-}=+1$), flat ($\alpha_{i}^{-}=0$), and dS
($\alpha_{i}^{-}=-1$) by setting parameters,
$\alpha_{o}^{-}=+1,\omega=+1$, $k_{o/i}^{-}=+1/2$, ${
m_{0}=1}$.}{fig:numpics5}

\figs{7cm}{7cm}{EflatVgpf}{7cm}{7cm}{EdsVgpf}{\small Plots of the
effective potentials for the shell with a linear fluid and the
exterior flat space ($\alpha^+_o=0,\alpha^+_i=+1,
k_{o}^{+}=+1,k^+_i=+1/4$: LHS)/dS space
($\alpha^+_{o/i}=+1,k_{o}^{+}= (x^+_C)^2 =+1,k^+_i=+1/4$: RHS). We
choose an interior AdS space with a point mass
($\alpha_{i}^{-}=+1$), flat ($\alpha_{i}^{-}=0$), and dS
($\alpha_{i}^{-}=-1$) by setting parameters,
$\alpha_{o}^{-}=+1,\omega=+1$, $k_{o/i}^{-}=+1/2$,
${m_{0}=4}$.}{fig:numpics4}

\subsection{Chaplygin-Gas Shells: $n, \omega <0$}

For the Chaplygin-type gas shells ($n,\omega <0$), by setting
$n=-1/(\lambda +1)$ one can conveniently rewrite the equation of
state\footnote{For a close connection to D-branes, see Ref.
\cite{hjp}.} as $p=-\hat{A}/\pi \rho^{\lambda}= - A (m_{0}/2 \pi^2
\ell)(m_{0}/2 \pi \ell \rho)^{\lambda}$ and $\omega =-\hat{A}
(m_{0}/2 \pi \ell)^{1/n}=-A$, in agreement with more standard
conventions \cite{chap}. Then, for finite $n$, i.e., $\lambda \neq
-1$ one finds
\begin{equation}
\label{rho:finite_-n} \rho(x) = \frac{m_{0}}{2 \pi \ell}
\left[A+(1-A) x^{-(\lambda+1)}\right]^{1/(\lambda+1)},
\end{equation}
from Eq. (\ref{rho:finite_n}).

The shape of the effective potential depends on several parameters
$A$, $\alpha_{o/i}^{\pm}$, and $k_{o/i}^{\pm}$,  as shown in Fig.
\ref{fig:numpics10}. Near $x=0$, the effective potential behaves as
\begin{eqnarray}
V_{{\rm eff}}(x) \approx \frac{1}{4} \frac{J^2/\ell^2}{x^2}
-\frac{m_{0}^2 (1-A)^2}{16 } - \frac{(\alpha_{i}^{+}
k_{o}^{+}+\alpha_{o}^{+} k_{i}^{+}-\alpha_{i}^{-}
k_{o}^{-}-\alpha_{o}^{-} k_{i}^{-})^{2}}{m_{0}^2 (1-A)^2}
\end{eqnarray}
due to $\rho(x) \approx (m_{0}/2 \pi \ell) (1-A)$ for $A \neq 1$,
from Eq. (\ref{rho:finite_-n}). Again there is a centrifugal barrier
near the origin. The asymptotic behaviour as $x\rightarrow \infty$
is given by
\begin{eqnarray}%
V_{{\rm eff}}(x) \approx \frac{1}{2}\left[ (\alpha_{o}^{+}
\alpha_{i}^{+}+\alpha_{o}^{-} \alpha_{i}^{-})  - \frac{1}{16}
m_{0}^2 A^{2/(\lambda+1)}\right] x^2
\end{eqnarray}
and this depends on the values of $A$ and $\alpha_{o/i}^{\pm}$,
similarly to Fig. \ref{fig:effp}. In addition, as
 can be seen from a consideration of the effective potential
(\ref{V_eff:p}), it will not diverge for any finite value of $x$
when $A<1$, while it will {\it negatively} diverge due to the
vanishing energy density, i.e., $\rho \rightarrow 0$ at $x=x_A
\equiv [(A-1)/A]^{1/(\lambda+1)}$ when $A>1$ for all $\lambda>-1$.

On the other hand, for $A=1$, we have a uniform density
$\rho(x)=m_{0}/2 \pi \ell$ from Eq. (\ref{rho:finite_-n}) also and
the (full) effective potential is given by
\begin{equation}
 V_{{\rm eff}}(x) = \frac{1}{2}({\mathcal
N}_{+}^2+{\mathcal N}_{-}^2)-\frac{1}{16}m_{0}^2 x^2
-\frac{1}{m_{0}^2 x^2}({\mathcal N}_{+}^2-{\mathcal N}_{-}^2)^2.
\end{equation}
Near $x=0$, this behaves as
\begin{eqnarray}
\label{V:bounce} V_{{\rm eff}}(x) \approx \frac{1}{4}
\frac{J^2/\ell^2}{x^2} - \frac{(\alpha_{i}^{+}
k_{o}^{+}+\alpha_{o}^{+} k_{i}^{+}-\alpha_{i}^{-}
k_{o}^{-}-\alpha_{o}^{-} k_{i}^{-})^2}{m_{0}^2x^2}
-\frac{m_{0}^2}{16 }
\end{eqnarray}
and it depends on initial data: One has an infinite well for $J^2<4
\ell^2 (\alpha_{i}^{+} k_{o}^{+}+\alpha_{o}^{+}
k_{i}^{+}-\alpha_{i}^{-} k_{o}^{-}-\alpha_{o}^{-} k_{i}^{-})^2/
m_{0}^2 $   (this is what has been plotted in Fig.
\ref{fig:numpics10}), and an infinite barrier for $J^2>4\ell^2
(\alpha_{i}^{+} k_{o}^{+}+\alpha_{o}^{+} k_{i}^{+}-\alpha_{i}^{-}
k_{o}^{-}-\alpha_{o}^{-} k_{i}^{-})^2/ m_{0}^2 $. The case
$J^2=4\ell^2 (\alpha_{i}^{+} k_{o}^{+}+\alpha_{o}^{+}
k_{i}^{+}-\alpha_{i}^{-} k_{o}^{-}-\alpha_{o}^{-} k_{i}^{-})^2/
m_{0}^2 $ is a marginal case that has a finite well  and a finite
intrinsic curvature; the intrinsic curvature is finite even at the
point $x=0$, from Eq. (\ref{eq:ricci})\footnote{  The only relevant
term is the $x^2$ term (omitted in (\ref{V:bounce})), which cancels
the $1/x$ factor in Eq. (\ref{eq:ricci}). }, and so $x=0$ is a
bounce point. As $x \rightarrow \infty$, the effective potential
behaves as
\begin{eqnarray}
V_{{\rm eff}}(x) \approx \left[ \frac{1}{2} (\alpha_{o}^{+}
\alpha_{i}^{+}+\alpha_{o}^{-}\alpha_{i}^{-}) -\frac{m_{0}^2}{16 }
-\frac{1}{m_{0}^2} (\alpha_{o}^{+}
\alpha_{i}^{+}-\alpha_{o}^{-}\alpha_{i}^{-})^2 \right] x^2
\end{eqnarray}
which depends on the initial data also.

As an explicit example, we consider $\lambda=1$, describing a
conventional Chaplygin gas shell \cite{chap}. Then the effective
potential becomes
\begin{equation}
V_{{\rm eff}}(x)=\frac{1}{m_{0}^2(Ax^2+1-A)x^2} (a_{8} x^6 + a_6 x^4
+ a_4 x^2 + a_2),
\end{equation}
where
\begin{eqnarray}
&&a_{8} = -\left[m_{0}^4 A^2/16  - (\alpha_{o}^{+}\alpha_{i}^{+}
+\alpha_{o}^{-}\alpha_{i}^{-})(m_{0}/2)^4 A/2+
(\alpha_{o}^{+}\alpha_{i}^{+}-\alpha_{o}^{-}\alpha_{i}^{-})^2\right], \nonumber\\
&&a_{6}= -m_{0}^4 A (1-A)/8 + \left[(\alpha_{o}^{+} k_{i}^{+} +
\alpha_{i}^{+} k_{o}^{+} +
\alpha_{i}^{-}k_{o}^{-}+\alpha_{o}^{-}k_{i}^{-})A +
(\alpha_{o}^{+}\alpha_{i}^{+}+\alpha_{o}^{-}\alpha_{i}^{-})(1-A)
\right]
m_{0}^2/2\nonumber\\
&&\qquad
-2(\alpha_{o}^{+}\alpha_{i}^{+}-\alpha_{o}^{-}\alpha_{i}^{-})
(k_{i}^{+}\alpha_{o}^{+}+\alpha_{i}^{+}k_{o}^{+} -
\alpha_{i}^{-}k_{o}^{-}-\alpha_{o}^{-}k_{i}^{-}
), \nonumber\\
&&a_{4} = -m_{0}^4 (1-A)^2/16 + \left[(\alpha_{o}^{+}k_{i}^{+} +
\alpha_{i}^{+}k_{o}^{+}
+\alpha_{o}^{-}k_{i}^{-}+\alpha_{i}^{-}k_{o}^{-} )(1-A)
+2k_{o}^{-}k_{i}^{-}A\right]
m_{0}^2/2\nonumber \\
&&\qquad -(\alpha_{o}^{+} k_{i} +\alpha_{i}^{+}k_{o}^{+}
+\alpha_{i}^{-}k_{o}^{-}+\alpha_{o}^{-}k_{i}^{-} )^2\nonumber\\
&&\qquad+ 4(\alpha_{o}^{+}\alpha_{o}^{-}k_{i}^{+}k_{i}^{-}
+\alpha_{i}^{+}\alpha_{i}^{-}k_{o}^{+}k_{o}^{-}
+ \alpha_{o}^{+}\alpha_{i}^{-}k_{i}^{+}k_{o}^{-}
+\alpha_{o}^{-}\alpha_{i}^{+}k_{o}^{+}k_{i}^{-}),\nonumber\\
&&a_{2}=m_{0}^2 k_{o}^{-}k_{i}^{-} (1-A).\nonumber
\end{eqnarray}
For a black hole spacetime outside the shell, the collapsing shell
may form a black hole within a finite time for $A \le 1$. But this
is not always the case for $A>1$ since there exists a singular point
at $x=x_{A}=\sqrt{(A-1)/A}$. If the horizon $x_H$ is located at
$x_{H} > x_{A}$, then it will form a black hole, while if
$x_{H}<x_{A}$, it will form a finite-sized ring singularity unless
the numerator vanishes at the point $x_A$, $a_8 x_A^6 +a_6 x_A^4 +
a_4 x_A^2 +a_2=0$. Apart from a contrivance of very restrictive
conditions on the parameters $m_{0}, \alpha^{\pm}_{o/i},
k^{\pm}_{o/i}$, and $A$, this is still a somewhat singular
configuration: even though the intrinsic Ricci scalar of the shell
is finite and its energy density vanishes, the pressure diverges at
the point $x_A$. This suggests a bounce solution.

Comparing to the non-rotating case \cite{mo}, we again find that
angular momenta  does not in general prevent the emergence of a
naked ring singularity at a finite position of $x=x_{A}$. Violation
of cosmic censorship occurs from the gravitational shell collapse,
regardless of the rotation and initial data.

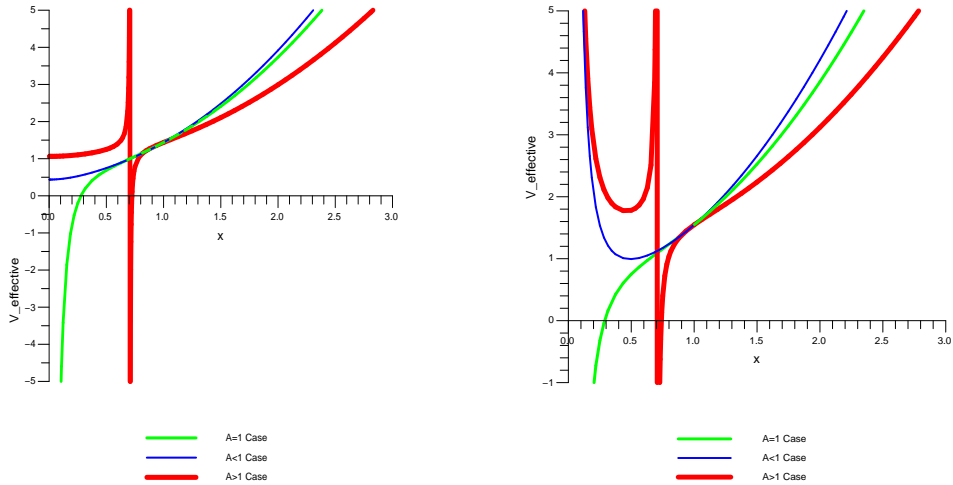
\figs{7cm}{7cm}{VgcgJ0}{7cm}{7cm}{Vgcgjn0}{\small Plots of the
effective potentials for the GCG shell with $\lambda=1$ and the
exterior/interior AdS spaces with point masses
($\alpha^{\pm}_{o/i}=+1, k_{o}^{+}=+1, k^+_i=+1/4,
k_{o/i}^{-}=+1/2, {m_{0}=2}$). (1) LHS: the case
of non-rotating shell (2) RHS: the case of rotating
shell.}{fig:numpics10}

\section{Discussion}\label{sec:discussion}

Our investigation of the gravitational collapse of rotating shells
in three dimensions has uncovered a number of interesting features.
We have studied whether or not angular momentum can significantly
change the collapse scenario and its resulting cosmic censorship
violations in the non-rotating cases in the literature.

For asymptotically AdS boundary conditions, we find that the
rotating shell collapses to either a black hole or a minimum value
and then expands out to infinity. For shells composed of
pressureless dust these are the only scenarios; the centrifugal
barrier forbids a naked singularity from forming.  However for
shells with pressure we have another scenario in which a naked
singular ring can form, violating cosmic censorship. When the
exterior spacetime is taken to be a  geometry with a point mass, one
might expect that the collapsing shell could form a curvature
singularity within finite time, since there is no event horizon as
with the non-rotating system in Ref. \cite{mo}. However we have
shown that a naked singularity never forms in the collapse of a
rotating dust shell due to a centrifugal barrier in the effective
potential experienced by the shell. Collapse scenarios for shells
with pressure show that a naked ring singularity of finite size can
be formed, where the effective potential and the surface
stress-energy tensor diverge. For asymptotically dS boundary
conditions a collapsing shell with pressure can form a naked
singularity, also. If the interior spacetime is (K)dS, then a
cosmological horizon can form from an expanding shell. Which of
these scenarios occurs depends on the choice of parameters and
initial conditions.

For asymptotically flat boundary conditions the qualitative
behaviours are similar, {\it i.e.}, a rotating collapsing dust shell
does not form a naked singularity, but a shell with pressure can
form a naked singularity.  For a polytropic shell this will be a
naked singular ring of finite size. The qualitative behaviour is
more or less intermediate between the asymptotically AdS and dS
cases as implied by the choice of $\alpha^{\pm}_{o/i}$ and
illustrated in Figs. \ref{fig:numpics2} and \ref{fig:numpics8},
though there are some anomalous regions that do not reveal this
simple trend.

The most intriguing lesson of this paper is that the centrifugal
barrier in the effective potential governing the time evolution of a
rotating dust shell can prevent formation of a naked singularity.
However if the shell has pressure, a ring singularity may form  for
a typical class of equations of state, i.e., $\omega >1$, where the
pressure dominates the angular momentum, with quite general initial
data. So if the exterior spacetime is assumed to be a geometry
without a (covered) black hole event horizon, then the singularity
may be naked and the violation of cosmic censorship is possible.

We have found that a collapsing shell  can form either a black
hole/cosmological horizon or a naked singularity or bounce to
infinity, depending on the initial data. This suggests a set of
phase transitions \cite{Pele:95}  along with accompanying critical
phenomena, similar to that discovered for scalar matter \cite{crit}.
It would be of great interest to study these phenomena explicitly as
they could reveal universal features of the critical exponents in
three dimensions.

Since we have confined our study to three dimensions (where there is
no gravitational radiation), some caution is warranted in applying
our results to higher dimensions. We have found that the angular
momentum does not in general prevent violation of cosmic censorship
in three dimensions, where local gravitational interactions vanish
outside of matter.  An interesting extension of our work would be to
include higher derivative terms, whose effect is to produce such
interactions. Another interesting extension is the inclusion of
quantum effects, since they might be expected to prevent singularity
formation, thereby sidestepping the cosmic censorship issue.  What
impact these modifications have on our results remains to be
investigated.

\vspace{5mm}
{\bf Acknowledgment}\\
{We would like to thank Sang Pyo Kim and Shin Nakamura for exciting
discussions and the headquarter of APCTP for warm hospitality during
the APCTP-TPI Joint Focus Program and Workshop. J. J. Oh would like
to thank Wontae Kim, Seungjoon Hyun, Hongbin Kim, Jaehoon Jeong,
Hyeong Chan Kim, Gungwon Kang, Inyong Cho, Constantinos
Papageorgakis, Andrew Strominger, and Chi-Ok Hwang for useful and
helpful discussions. R. B. Mann was supported by the Natural
Sciences and Engineering Research Council of Canada. J. J. Oh was
supported by the Korea Research Council of Fundamental Science \&
Technology (KRCF). M.-I. Park was supported by the Korea Research
Foundation Grant funded by Korea Government(MOEHRD)
(KRF-2007-359-C00011). }

\nc{\PR}[3]{Phys. Rev. {\bf #1}, #2 (#3)}
\nc{\NPB}[3]{Nucl. Phys. {\bf B#1}, #2 (#3)}
\nc{\PLB}[3]{Phys. Lett. {\bf B#1}, #2 (#3)}
\nc{\PRD}[3]{Phys. Rev. {\bf D#1}, #2 (#3)}
\nc{\PRL}[3]{Phys. Rev. Lett. {\bf #1}, #2 (#3)}
\nc{\PREP}[3]{Phys. Rep. {\bf #1}, #2 (#3)}
\nc{\EPJ}[3]{Eur. Phys. J. {\bf #1}, #2 (#3)}
\nc{\PTP}[3]{Prog. Theor. Phys. {\bf #1}, #2 (#3)}
\nc{\CMP}[3]{Comm. Math. Phys. {\bf #1}, #2 (#3)}
\nc{\MPLA}[3]{Mod. Phys. Lett. {\bf A #1}, #2 (#3)}
\nc{\CQG}[3]{Class. Quant. Grav. {\bf #1}, #2 (#3)}
\nc{\NCB}[3]{Nuovo Cimento {\bf B#1}, #2 (#3)}
\nc{\ANNP}[3]{Ann. Phys. (N.Y.) {\bf #1}, #2 (#3)}
\nc{\GRG}[3]{Gen. Rel. Grav. {\bf #1}, #2 (#3)}
\nc{\MNRAS}[3]{Mon. Not. Roy. Astron. Soc. {\bf #1}, #2 (#3)}
\nc{\JHEP}[3]{JHEP {\bf #1}, #2 (#3)}
\nc{\JCAP}[3]{JCAP {\bf #1},, #2 {(#3)}}
\nc{\ATMP}[3]{Adv. Theor. Math. Phys. {\bf #1}, #2 (#3)}
\nc{\AJP}[3]{Am. J. Phys. {\bf #1}, #2 (#3)}
\nc{\ibid}[3]{{\it ibid.} {\bf #1}, #2 (#3)}
\nc{\ZP}[3]{Z. Physik {\bf #1}, #2 (#3)}
\nc{\PRSL}[3]{Proc. Roy. Soc. Lond. {\bf A#1}, #2 (#3)}
\nc{\LMP}[3]{Lett. Math. Phys. {\bf #1}, #2 (#3)}
\nc{\AM}[3]{Ann. Math. {\bf #1}, #2 (#3)}
\nc{\hepth}[1]{{\tt [arxiv:hep-th/{#1},]}}
\nc{\grqc}[1]{{\tt [arxiv:gr-qc/{#1},]}}
\nc{\astro}[1]{{\tt [arxiv:astro-ph/{#1},]}}
\nc{\hepph}[1]{{\tt [arxiv:hep-ph/{#1},]}}
\nc{\phys}[1]{{\tt [arxiv:physics/{#1},]}}

\end{document}